\documentclass[twocolumn,aps,prd,amssymb,nofootinbib,eqsecnum]{revtex4}
\usepackage{epsfig}
\usepackage{amsmath}
\usepackage{amssymb}

\begin{document}
\title{Higher order gravity theories and scalar tensor theories}

\author{\'Eanna \'E.\ Flanagan}
\affiliation{Cornell University, Newman Laboratory, Ithaca, NY
14853-5001.}

\begin{abstract}
We generalize the known equivalence between higher order gravity
theories and scalar tensor theories to a new class of theories.  
Specifically, in the context of a first order or Palatini variational
principle where the metric and connection are treated as independent
variables, we consider theories for which the Lagrangian density is a
function $f$ of (i) the Ricci scalar 
computed from the metric, and (ii) a second Ricci scalar computed from the
connection.  We show that such theories can be written as 
tensor-multi-scalar theories with two scalar fields with the following
features:  (i) the two dimensional $\sigma$-model metric that defines
the kinetic energy terms for the scalar fields has constant, negative
curvature; (ii) the coupling function determining the coupling to matter
of the scalar fields is universal, independent of the choice of
function $f$; and (iii) if both mass eigenstates are long ranged, then the
Eddington post-Newtonian parameter $\gamma$ has value $1/2$.  Therefore
in order to be compatible with solar system experiments at least one
of the mass eigenstates must be short ranged. 
\end{abstract}

\maketitle


\section{Introduction and Summary}
\label{intro}

A well-known class of theories of gravity that modify general relativity
can be obtained
by taking the Lagrangian density to be some nonlinear
function $f$ of the Ricci scalar.  The action for such theories is 
\begin{equation}
S[{\bar g}_{\mu\nu},\psi_{\rm m}] = \frac{1}{2 \kappa^2} \int d^4 x
\sqrt{- {\bar g}} f({\bar R}) + S_{\rm m}[{\bar g}_{\mu\nu},\psi_{\rm
    m}],
\label{eq:action0}
\end{equation}
where ${\bar g}_{\mu\nu}$ is the metric (which we
have barred for later notational convenience), ${\bar R}$ is its Ricci
scalar, $f$ is the function, and $\kappa^2 = 8 \pi G$.  The second
term is the matter action $S_{\rm m}$, which is some functional of the
matter fields $\psi_{\rm m}$ and of the metric ${\bar g}_{\mu\nu}$
\footnote{We use units in which $\hbar=c=1$, and we use the sign
conventions of Ref.\ \protect{\cite{MTW}}.}.  The theory
(\ref{eq:action0}) is equivalent to \footnote{The equivalence was first
  shown, in the context of a 
specific choice of $f$, by Higgs \cite{higgs}, and later
independently by several different researchers
including Bicknell \cite{bicknell}, Teyssandier and Tourrenc \cite{TT},
and Whitt \cite{whitt}.  The generalization to
arbitrary functions $f$ was first given by Schmidt \cite{schmidt}.
Subsequently Wands \cite{Wands} once again independently discovered the
equivalence, and independently derived the result for arbitrary $f$.} a
scalar-tensor gravity theory \cite{Damour92}.

A second class of theories of gravity is given by re-interpreting the
action (\ref{eq:action0}) in terms of a Palatini or first-order
variational principle \cite{Vollick}.  In this reinterpretation, one
considers the action to be a functional of the matter fields $\psi_{\rm
m}$, a metric ${\bar g}_{\mu\nu}$, and a connection ${\hat
\nabla}_\mu$ which is independent of the metric.  The resulting
action is 
\begin{equation}
S[{\bar g}_{\mu\nu},{\hat \nabla}_\mu,\psi_{\rm m}] = \frac{1}{2
  \kappa^2} \int d^4 x \sqrt{- {\bar g}} f({\hat R}) +
S_{\rm m}[{\bar g}_{\mu\nu},\psi_{\rm m}],
\label{eq:action1}
\end{equation}
where ${\hat R} = {\bar g}^{\mu\nu} \,
{\hat R}_{\mu\nu}$ and ${\hat R}_{\mu\nu}$ is the Ricci
tensor of the connection ${\hat \nabla}_\mu$.
When $f({\hat R}) = {\hat R}$ (the case of general relativity), the
action (\ref{eq:action1}) gives rise to the same equations of motion
as the theory (\ref{eq:action0}) based on the standard or second order
variational principle.  However, for more general functions $f({\hat
  R})$, the theories (\ref{eq:action0}) and (\ref{eq:action1}) are not
equivalent \cite{Vollick,Ferraris}.

Modifications of gravity of both types (\ref{eq:action0}) and
(\ref{eq:action1}) have been suggested
\cite{Capo,Carroll,Lue,Vollick,Meng,Odint1}
to explain the observed recent acceleration of the Universe's
expansion \cite{SN,wmap}.  However,  there are as yet no successful
models along these lines.  Theories of the type (\ref{eq:action0})
contain a scalar field which couples to matter with the same strength
as does gravity, as pointed out by Chiba \cite{Chiba}.  These theories 
are thus ruled out by solar system experiments\footnote{It may be
possible to evade this constraint for certain choices of the function
$f({\bar R})$.  For example, for $f({\bar R}) \propto {\bar R}^2 (\ln
{\bar R})^n$, the potential $V(\Phi)$ for the scalar field is of the
form $V(\Phi) \propto 1/\Phi^n$ for large $\Phi$. For this potential,
and for the matter coupling for $\Phi$ that follows from the action
(\ref{eq:action0}), 
there exists a regime in which the coupling of the scalar field to
large objects is suppressed by nonlinear effects \cite{chameleon}.
However, it is not yet clear if this class of ``chameleon field''
theories give acceptable cosmologies.} unless the scalar field
is short ranged, and in that case it is difficult to obtain an
accelerating Universe \cite{Chiba}.  Theories of the type
(\ref{eq:action1}) are actually not
modifications of gravity at all \footnote{This statement is valid
classically.  However quantum mechanically things may be more
complicated.}, despite appearances.  Instead, these 
theories, when written in a canonical form, contain a scalar field
with no kinetic energy energy term.  When one integrates out the
scalar field one obtains general relativity coupled to a modified
matter action, not the original matter action $S_{\rm m}[{\bar
g}_{\mu\nu},\psi_{\rm m}]$ that appears in Eq.\
(\ref{eq:action1}) \cite{Flanagan}.

This paper is based on the observation, due to Nima Arkani-Hamed
\cite{Nima}, that theories of the form (\ref{eq:action1}) are unstable 
to matter loop corrections.  Specifically, matter loops will give rise 
to a correction to the action (\ref{eq:action1}) proportional to the
Ricci scalar ${\bar R}$ of the metric
${\bar g}_{\mu\nu}$.  In general relativity
these corrections merely renormalize Newton's constant of gravitation.
Here, however, such corrections give rise to a Lagrangian density for
the gravitational action which is no longer a function of ${\hat R}$
alone, but instead is a function of both ${\hat R}$ and ${\bar R}$,
the Ricci scalars of the connection and of the metric.
Thus, the form of the gravitational action (\ref{eq:action1}) is not
preserved under loop corrections, and there is a fine-tuning problem
inherent in assuming an action of the form (\ref{eq:action1}).  

That fine tuning problem is also apparent in the alternative description of
the theory (\ref{eq:action1}), derived in Ref.\ \cite{Flanagan}, as a
type of scalar-tensor type theory without any kinetic energy term for the
scalar field \cite{Nima}.  The action of that alternative description is
\begin{eqnarray}
S[g_{\mu\nu},\Phi,\psi_{\rm m}] &=& 
\int d^4 x
\sqrt{- g} \bigg[ \frac{R}{2 \kappa^2}  - V(\Phi)
\bigg] 
\nonumber \\ && 
+ S_{\rm m}[e^{2 \alpha(\Phi)} g_{\mu\nu},\psi_{\rm m}],
\label{eq:action2}
\end{eqnarray}
where $\Phi$ is the auxiliary scalar field with coupling function
$\alpha(\Phi)$ and potential $V(\Phi)$, and 
$g_{\mu\nu} = \exp[-2 \alpha(\Phi)] {\bar g}_{\mu\nu}$ is the
Einstein frame metric.  In terms of these variables, the correction
term due to matter loops is proportional to ${\bar R} = e^{-2 \alpha}
[R - 6 ({\bar \nabla} \alpha)^2 - 6 {\bar \Box} \alpha ]$.  Therefore  
matter loops induce a kinetic term for the field $\Phi$, which becomes
a dynamical field, and the loop-corrected theory is a genuine
modification of general relativity, unlike the original theory
(\ref{eq:action1}) or (\ref{eq:action2}).

These considerations suggest that, in the context of the first-order
variation formalism, it is natural to consider a more general class of
theories of gravity of the form
\begin{equation}
S[{\bar g}_{\mu\nu},{\hat \nabla}_\mu,\psi_{\rm m}] = \frac{1}{2
  \kappa^2} \int d^4 x \sqrt{- {\bar g}} f({\bar R},{\hat R}) +
S_{\rm m}[{\bar g}_{\mu\nu},\psi_{\rm m}].
\label{eq:action3}
\end{equation}
Here the Lagrangian density of the gravitational action
is some function $f$ of both the Ricci
scalar ${\bar R}$ of the metric ${\bar g}_{\mu\nu}$ and the Ricci
scalar ${\hat R}$ of the connection ${\hat \nabla}_{\mu}$.  
These theories are stable under loop corrections, which
just give rise to corrections to the function $f({\bar R}, {\hat R})$.
The purpose of this paper is to analyze the class of theories
(\ref{eq:action3}), and to derive for these theories an equivalent
description, at the classical level, as tensor-biscalar theories.

\subsection{Summary of results}

We now turn to a description of our results.  
We define
\begin{equation}
f_1({\bar R},{\hat R}) = \frac{\partial f}{\partial {\bar R}}({\bar
  R},{\hat R}),\ \ \ \ \ \ 
f_2({\bar R},{\hat R}) = \frac{\partial f}{\partial {\hat R}}({\bar
  R},{\hat R}),
\end{equation}  
and we define the signs
\begin{equation}
\varepsilon_1 = {\rm sign} f_1,\ \ \ \ \ 
\varepsilon_2 = {\rm sign} f_2,
\label{eq:varepsilon12def}
\end{equation}
and
\begin{equation}
\varepsilon_3 = {\rm sign} (f_1 + f_2).
\label{eq:varepsilon3def}
\end{equation}
Our construction in Sec.\ \ref{sec:derivation} below shows that 
if $\varepsilon_1$, $\varepsilon_2$ and $\varepsilon_3$
are constant in an initial data set, then they will be constant
throughout the corresponding future Cauchy evolution.  We restrict
attention to this portion of phase space, in which the signs
$\varepsilon_1$, $\varepsilon_2$ and $\varepsilon_3$ are constants,
independent of space and time.   The reason for this restriction is that
in regions of phase
space where one or more of these variables flips sign somewhere in
spacetime, we suspect that the theory does not posses a well posed
initial value 
formulation \footnote{The reason for this suspicion is that points in
spacetime of at which $\varepsilon_1$, $\varepsilon_2$ or $\varepsilon_3$
flip sign are points at which one of the three kinetic terms in
the action (\ref{eq:action7}) below vanishes.  The initial value
formulation is known to break down in 
other, simpler contexts in which this type of phenomenon occurs.
For example, the theory $S = \int d^4 x \sqrt{-g} [R/(2 \kappa^2)
  - (\nabla \Phi)^2/2 - \xi R \Phi^2 ]$ has a good initial value
formulation in the region of phase space $\Phi \le \Phi_c$, where
$\Phi_c = 1 / \sqrt{2 \kappa^2 \xi}$, but not once $\Phi$ crosses the
value $\Phi_c$, since $\Phi \equiv \Phi_c$, $g_{\mu\nu}=$ (anything)
is a solution of the equations of motion.}.
Assuming that $\varepsilon_1$,
$\varepsilon_2$, and $\varepsilon_3$ are constants, we show below that 
in order for the theory not to contain ghosts or negative energy
excitations, we must have
\begin{equation}
\varepsilon_1 = 1,
\label{eq:stability1a}
\end{equation}
\begin{equation}
\varepsilon_2 = -1,
\label{eq:stability1b}
\end{equation}
and
\begin{equation}
\varepsilon_3 = 1.
\label{eq:stability2}
\end{equation}

Next, we further restrict the region of phase space under
consideration, and the set of allowed functions $f$,
as follows.  We
define $\Delta({\bar R},{\hat R})$ to be
the determinant of the matrix of second order partial derivatives of
$f$ with respect to ${\bar R}$ and ${\hat R}$.  We assume that there exists
an open domain ${\cal D}$ of points
$({\bar R}, {\hat R})$ for which the following three conditions are
satisfied: (i) We have
\begin{equation}
\Delta({\bar R}, {\hat R}) \ne 0
\label{eq:assumption0}
\end{equation}
for all points $({\bar R},{\hat R})$ in ${\cal D}$;
(ii) The conditions (\ref{eq:stability1a}) -- (\ref{eq:stability2})
 are satisfied for all points $({\bar R},{\hat R})$ in ${\cal D}$; and
(iii) Equations (\ref{eq:transl1}) and (\ref{eq:transl2_I}) below
define a bijection between ${\cal D}$ and some open domain ${\cal
  D}^\prime$ of field values $(\Phi,\Psi)$.
Then, the theory (\ref{eq:action3}) with initial data in the domain
${\cal D}$ is equivalent to a tensor-biscalar theory with field values
in the domain ${\cal D}^\prime$. 

We next describe this tensor-biscalar theory.
The action for a general
tensor-multi-scalar theory is the following functional of the matter
fields $\psi_{\rm m}$, the Einstein-frame metric $g_{\mu\nu}$, and a
N-tuple of scalar fields $\Phi^A = (\Phi^1, \ldots , \Phi^N)$ \cite{Damour92}:
\begin{eqnarray}
S[g_{\mu\nu},\Phi^A,\psi_{\rm m}] &=&
\int d^4 x
\sqrt{- g} \bigg[ \frac{R}{2 \kappa^2}  -  V(\Phi^A)
\nonumber \\ && 
- \frac{1}{2}
  \gamma_{AB}(\Phi^C) g^{\mu\nu} \nabla_\mu \Phi^A \nabla_\nu \Phi^B 
\bigg] 
\nonumber \\ && 
+ S_{\rm m}[e^{2 \alpha(\Phi^A)} g_{\mu\nu},\psi_{\rm m}].
\label{eq:action4}
\end{eqnarray}
This action is characterized by the coupling function $\alpha(\Phi^A)$
and potential $V(\Phi^A)$ of the scalar fields, and by the
$\sigma$-model metric $\gamma_{AB}(\Phi^C)$.  For the theory
(\ref{eq:action3}) we have 
$N=2$, $\Phi^A = (\Phi^1,\Phi^2) =
(\Phi,\Psi)$, the $\sigma$-model metric is given by
\begin{equation}
\gamma_{AB}(\Phi^C) d\Phi^A d\Phi^B = d\Phi^2 + \cosh^2 \left(\frac{\kappa}
{\sqrt{6}} \Phi \right) d\Psi^2,
\label{eq:sigmametricI}
\end{equation}
and the coupling function is
\begin{equation}
\alpha(\Phi,\Psi) = \frac{\kappa}{\sqrt{6}} \Psi + \ln \cosh \left(
\frac{\kappa}{\sqrt{6}} \Phi \right).
\label{eq:alphadefI}
\end{equation}
The potential is given by the formula
\begin{equation}
V(\Phi,\Psi) = \frac{e^{4 \alpha(\Phi,\Psi)} }{2 \kappa^2} 
\left[
  \varphi f_1(\varphi,\psi) + \psi f_2(\varphi,\psi) - f(\varphi,\psi) \right].
\label{eq:Vdef}
\end{equation}
Here we have introduced two additional scalar fields $\varphi$ and
$\psi$, which are given in terms of $\Phi$ and $\Psi$ by the equations
\footnote{The left hand sides of Eq.\ (\ref{eq:transl2_I}) 
is less than one by Eqs.\
(\ref{eq:varepsilon3def}) and (\ref{eq:stability2}), so it is
possible to solve this equation for $\Phi$.} 
\begin{equation}
 \left| f_1(\varphi,\psi) \right| = 
\exp \left[ - 2 \frac{\kappa}{\sqrt{6}} \Psi \right],
\label{eq:transl1}
\end{equation}
and
\begin{equation}
\frac{  \left| f_2(\varphi,\psi) \right| }{ \left| f_1(\varphi,\psi)
  \right| } = \tanh^2 \left[ \frac{\kappa}{\sqrt{6}} \Phi \right].
\label{eq:transl2_I}
\end{equation}

For specific choices of the function $f({\bar R},{\hat R})$ it is
straightforward to compute the potential $V$ from Eqs.\ (\ref{eq:Vdef}) --
(\ref{eq:transl2_I}).  For example, if $f = {\bar R} - {\hat R} +
{\hat R} {\bar R} / m^2$ we have $V = m^2 u (1-u) / (2 \kappa^2)$,
where
\begin{equation}
u = \cosh^2 \left( \frac{\kappa}{\sqrt{6}} \Phi \right) \left[ 1 -
\exp \left( 2 \frac{\kappa}{\sqrt{6}} \Psi \right) \right].
\end{equation}

\section{Derivation of equivalence}
\label{sec:derivation}

Consider the gravitation theory given by the action
(\ref{eq:action3}).  An action which is equivalent to this at the
classical level can be obtained by using the technique of Refs.\
\cite{TT,Wands} of introducing auxiliary scalar field(s).
Specifically, we introduce two scalar fields $\varphi$ and $\psi$, and
we define the action
\begin{eqnarray}
&& {\tilde S}[{\bar g}_{\mu\nu},{\hat \nabla}_\mu,\varphi,\psi,\psi_{\rm m}] = \frac{1}{2
  \kappa^2} \int d^4 x \sqrt{- {\bar g}} \bigg[ f(\varphi,\psi) 
\nonumber \\ &&
+ ({\bar R} - \varphi) f_1(\varphi,\psi) + ({\hat R}-\psi)
  f_2(\varphi,\psi) \bigg] 
\nonumber \\ &&+ 
S_{\rm m}[{\bar g}_{\mu\nu},\psi_{\rm m}].
\label{eq:action5}
\end{eqnarray}
Then the equations of motion for $\varphi$ and $\psi$ enforce $\varphi
= {\bar R}$, $\psi = {\hat R}$, as long as 
\begin{equation}
{\rm det} \ 
\left[
\begin{array}{cc}
\frac{\partial^2 f}{\partial \varphi^2}
&  \frac{\partial^2 f}{\partial \varphi \partial \psi} \\
\frac{\partial^2 f}{\partial \varphi \partial \psi} 
&  \frac{\partial^2 f}{\partial \psi^2}
\end{array}\right] \ne 0.
\label{eq:condt1}
\end{equation}
The condition (\ref{eq:condt1}) will always be satisfied because of
our assumption (\ref{eq:assumption0}) above.  Therefore the
actions (\ref{eq:action3}) and (\ref{eq:action5}) are classically
equivalent.  

We write the action (\ref{eq:action5}) as 
${\tilde S} = {\tilde S}_1 + {\tilde S}_2 + {\tilde S}_3 + S_{\rm m}$,
where
\begin{equation}
{\tilde S}_1 = \frac{1}{2 \kappa^2} \int d^4x \sqrt{- {\bar g}} {\bar
  R} f_1(\varphi,\psi),
\end{equation}
\begin{equation}
{\tilde S}_2 = \frac{1}{2 \kappa^2} \int d^4x \sqrt{- {\bar g}} {\hat
  R} f_2(\varphi,\psi),
\end{equation}
and
\begin{equation}
{\tilde S}_3 = \frac{1}{2 \kappa^2} \int d^4x \sqrt{- {\bar g}} 
\left[ f - \varphi f_1 - \psi f_2  \right].
\label{eq:actionS3}
\end{equation}
The action ${\tilde S}_1$ can be rewritten as an Einstein-frame
form using the following standard procedure.
We define the scalar field $\rho$ by 
\begin{equation}
f_1(\varphi,\psi) = \varepsilon_1 e^{-2 \rho},
\label{eq:rhodef}
\end{equation}
where $\varepsilon_1 = \pm 1$ [{\it cf}.\ Eq.\
  (\ref{eq:varepsilon12def}) above]. 
Defining the conformally transformed metric 
\begin{equation}
{\tilde g}_{\mu\nu} =
  \exp[- 2 \rho] {\bar g}_{\mu\nu},
\label{eq:tildegdef}
\end{equation}
and using ${\bar R} = e^{-2 \rho}
[{\tilde R} - 6 ({\tilde \nabla} \rho)^2 - 6 {\tilde \Box} \rho ]$,
we obtain
\begin{equation}
{\tilde S}_1[{\tilde g}_{\mu\nu},\varphi,\psi] =
\frac{\varepsilon_1}{2 \kappa^2} \int d^4 x \sqrt{- {\tilde g}} 
\left[ {\tilde R} - 6 ( {\tilde \nabla} \rho)^2 \right].
\label{eq:actionS1}
\end{equation}

A similar transformation can be carried out for the action ${\tilde
S}_2$.  We define the scalar field $\sigma$ by 
\begin{equation}
f_2(\varphi,\psi) = \varepsilon_2 e^{-2 \sigma},
\label{eq:sigmadef}
\end{equation}
where $\varepsilon_2 = \pm 1$ [{\it cf}.\ Eq.\
  (\ref{eq:varepsilon12def}) above].  
Now the Ricci scalar ${\hat R}$ that appears in the action
${\tilde S}_2$ depends both on the metric ${\bar g}_{\mu\nu}$ and on
the connection ${\hat \nabla}_\mu$ via
\begin{equation}
{\hat R} = {\bar g}^{\mu\nu} {\hat R}_{\mu\nu},
\label{eq:hatRdef}
\end{equation}
where ${\hat R}_{\mu\nu}$ is the Ricci tensor of the connection ${\hat
\nabla}_\mu$.  Under conformal transformations of the metric
that preserve the connection, this quantity will undergo a simple,
multiplicative transformation coming from the first factor in Eq.\
(\ref{eq:hatRdef}).  Thus, if we define 
\begin{equation}
{\check g}_{\mu\nu} =
  \exp[- 2 \sigma] {\bar g}_{\mu\nu}
\label{eq:checkgdef}
\end{equation}
we obtain 
\begin{equation}
{\tilde S}_2[{\check g}_{\mu\nu},{\hat \nabla}_\mu,\varphi,\psi] = \frac{\varepsilon_2}{2 \kappa^2} \int d^4 x \sqrt{- {\check g}}
( {\check g}^{\mu\nu} {\hat R}_{\mu\nu}).
\label{eq:actionS2}
\end{equation}

Next, we note that the total action
(\ref{eq:action5}) depends on the connection ${\hat \nabla}_\mu$ only through
the term (\ref{eq:actionS2}).  Also the action (\ref{eq:actionS2}) is
just the Palatini-variation version of the Einstein-Hilbert action of
general relativity.  Therefore, the equation of motion for ${\hat \nabla}_\mu$
stipulates that ${\hat \nabla}_\mu$ is the connection ${\check
  \nabla}_\mu$ determined by the metric ${\check g}_{\mu\nu}$.  
In the classical theory, we are free to substitute the relation 
${\hat \nabla}_\mu = {\check \nabla}_\mu$ back
into the action (\ref{eq:actionS2}) to obtain
\begin{equation}
{\tilde S}^\prime_2[{\check g}_{\mu\nu},\varphi,\psi] =
\frac{\varepsilon_2}{2 \kappa^2} \int d^4 x \sqrt{- {\check g}}
{\check R},
\label{eq:actionS2p}
\end{equation}
where ${\check R}$ is the Ricci scalar of the metric ${\check
g}_{\mu\nu}$.  In other words, we obtain an equivalent 
theory at the classical level by replacing the Palatini-variation action
${\tilde S}_2$ with its standard-variation version 
${\tilde S}^\prime_2$.

Next, we write the action (\ref{eq:actionS2p}) in terms of the metric
${\tilde g}_{\mu\nu}$ using the definitions (\ref{eq:tildegdef}) and
(\ref{eq:checkgdef}).  This gives
\begin{equation}
{\tilde S}_2^\prime[{\tilde g}_{\mu\nu},\varphi,\psi] = 
\frac{\varepsilon_2}{2 \kappa^2} \int d^4 x \sqrt{- {\tilde g}} 
\left[ e^{2 \tau} {\tilde R} + 6 e^{2 \tau} ( {\tilde \nabla} \tau)^2 \right],
\end{equation}
where
\begin{equation}
\tau = \rho - \sigma.
\label{eq:taudef}
\end{equation}
Adding this to the action (\ref{eq:actionS1}) and rewriting in terms
of the conformally transformed metric
\begin{equation}
g_{\mu\nu} = e^{- 2 \chi} {\tilde g}_{\mu\nu}
\label{eq:gdef}
\end{equation}
gives
\begin{eqnarray}
{\tilde S}_1 + {\tilde S}_2^\prime &=& 
\frac{1}{2 \kappa^2} \int d^4 x \sqrt{- g} 
\bigg\{ 6 \varepsilon_2 e^{2 \chi} e^{2 \tau} (\nabla \tau)^2 
\nonumber \\ && 
+e^{2 \chi} (\varepsilon_1 + \varepsilon_2 e^{2 \tau}) 
\left[ R - 6 (\nabla \chi)^2 - 6 \Box \chi \right]
\nonumber \\ &&
- 6 \varepsilon_1 e^{2 \chi} (\nabla \rho)^2 \bigg\}.
\label{eq:action6}
\end{eqnarray}

Next, it follows from the definitions (\ref{eq:rhodef}),
(\ref{eq:sigmadef}) and (\ref{eq:taudef}) of $\rho$, $\sigma$ and
$\tau$ and the definition (\ref{eq:stability2}) of $\varepsilon_3$
that 
\begin{equation}
\varepsilon_3 = {\rm sign} (\varepsilon_1 + \varepsilon_2 e^{2 \tau}).
\label{eq:varepsilon3formula}
\end{equation}
We now choose the conformal
factor $e^{2 \chi}$ entering into the definition (\ref{eq:gdef}) of the
metric $g_{\mu\nu}$ to satisfy
\begin{equation}
e^{2 \chi} | \varepsilon_1 + \varepsilon_2 e^{2 \tau} | =1.
\label{eq:chidef}
\end{equation}
Using Eqs.\ (\ref{eq:varepsilon3formula}) and (\ref{eq:chidef}) the
action (\ref{eq:action6}) simplifies to 
\begin{eqnarray}
{\tilde S}_1 + {\tilde S}_2^\prime &=& 
\frac{1}{2 \kappa^2} \int d^4 x \sqrt{- g} 
\bigg[ \varepsilon_3 R - \frac{6 \varepsilon_1
    \varepsilon_3}{\varepsilon_1 +     \varepsilon_2 e^{2 
    \tau}} (\nabla \rho)^2
\nonumber \\ && 
+  \frac{6 \varepsilon_1 \varepsilon_2 \varepsilon_3 e^{2 \tau}}{ 
(\varepsilon_1 + \varepsilon_2 e^{2 \tau})^2}
(\nabla \tau)^2 \bigg].
\label{eq:action7}
\end{eqnarray}

Now the sign of the coefficient of the Ricci scalar $R$ in the action
(\ref{eq:action7}) must be positive, otherwise the matter action
$S_{\rm m}$ will contain fields whose kinetic terms have 
the wrong sign relative to the gravitational action, giving rise to an
unstable theory.  This yields the
condition (\ref{eq:stability2}).
Similarly, requiring that the last two terms in the action
(\ref{eq:action7}) have their 
conventional signs
\footnote{If the 
sign of a scalar-field action relative to the gravitational action is
flipped from   
its usual value, then the Hamiltonian of the theory is unbounded below
and one expects the theory to be quantum mechanically unstable.  Note
however that in such theories the decay timescales may be long enough
for the theory to make sense as an effective field theory over a
certain range of scales, as suggested in Refs.\
\protect{\cite{Damour92,Carroll2}}.  In this paper we restrict
attention to sectors where the signs of the scalar-field 
actions have the conventional value.} and using the relation
(\ref{eq:varepsilon3formula}) yields the conditions
(\ref{eq:stability1a}) and (\ref{eq:stability1b}).

Next, we define the fields $\Phi$ and $\Psi$ by
\begin{equation}
\Phi = \frac{\sqrt{6}}{\kappa} {\rm
  sign}(\tau) \tanh^{-1} \left[ e^{- 
    |\tau| } \right]
\label{eq:Phidef}
\end{equation}
and
\begin{equation}
\Psi = \frac{\sqrt{6}}{\kappa} \rho.
\label{eq:Psidef}
\end{equation}
It follows that 
\begin{equation}
\frac{1}{\varepsilon_1 + \varepsilon_2 e^{2 \tau}}
= \cosh^2(\kappa \Phi/\sqrt{6}),
\label{eq:useful}
\end{equation}
and substituting into the action (\ref{eq:action7}) reproduces the
first and third terms of the action (\ref{eq:action4}).
The last term in the action (\ref{eq:action4}) is just the fourth term
in the action (\ref{eq:action5}) where we have used the relation
\begin{equation}
{\bar g}_{\mu\nu} = e^{2 \alpha} g_{\mu\nu}
\label{eq:alphadef}
\end{equation}
between the Jordan-frame metric ${\bar g}_{\mu\nu}$ and the Einstein
frame metric $g_{\mu\nu}$.  In order to compute the coupling function
$\alpha$ that is defined by the relation (\ref{eq:alphadef}), we
combine the definitions (\ref{eq:tildegdef}) and (\ref{eq:gdef}) of
${\tilde g}_{\mu\nu}$ and $g_{\mu\nu}$ together with the formula
(\ref{eq:chidef}) for the conformal factor $e^{2 \chi}$. 
This yields
\begin{equation}
e^{2 \alpha} = \frac{e^{2 \rho}}{\varepsilon_1 + \varepsilon_2 e^{2
    \tau}},
\end{equation}
and combining this with Eqs.\ (\ref{eq:Psidef}) and (\ref{eq:useful})
yields the formula (\ref{eq:alphadefI}).  Finally, the expression
(\ref{eq:Vdef}) for the potential is obtained by combining Eqs.\
(\ref{eq:actionS3}) and (\ref{eq:alphadef}), and the formulae
(\ref{eq:transl1}) 
and (\ref{eq:transl2_I}) for $\varphi$ and $\psi$ in terms of $\Phi$ and $\Psi$
can be obtained from the definitions (\ref{eq:rhodef}),
(\ref{eq:sigmadef}), (\ref{eq:taudef}), (\ref{eq:Phidef}) and
(\ref{eq:Psidef}).

Finally, we note that the condition (\ref{eq:stability2}) 
together with
Eqs.\ (\ref{eq:stability1a}), (\ref{eq:stability1b}) 
and (\ref{eq:varepsilon3formula}) imply that
$\tau \le 0$, which implies from Eq.\ (\ref{eq:Phidef}) that 
\begin{equation}
\Phi \le 0.
\label{eq:restriction}
\end{equation}
In other words, the two theories are equivalent in the following
sense: solutions of the original theory (\ref{eq:action3})
can be mapped onto solutions of the tensor-biscalar theory (\ref{eq:action4})
that satisfy (\ref{eq:restriction}).  This restriction is somewhat
puzzling as the 
condition (\ref{eq:restriction}) is clearly not preserved under general
dynamical evolution.  The resolution of this puzzle is that if
one starts with an initial data set satisfying (\ref{eq:restriction})
and evolves forward using the tensor-biscalar theory until the
condition (\ref{eq:restriction}) is violated, the locus of
points at which $\Phi=0$ corresponds to a singularity of the
original variables ${\bar g}_{\mu\nu}$, ${\hat \nabla}_\mu$.  
This can be seen from Eq.\ (\ref{eq:transl2_I}), since we have argued
that $\varepsilon_2 = {\rm sign}(f_2)$ cannot flip sign, so $f_1$ must pass
through $\pm \infty$ as $\Phi$ passes through $0$.

We also note that the methods of this paper can be applied to special
choices of the function $f({\bar R},{\hat R})$ for which the quantity
$\Delta({\bar R},{\hat R})$ vanishes identically, {\it cf.} Eq.\ 
(\ref{eq:assumption0}) above.  In such cases one
obtains a theory with one scalar field.  The choice $f({\bar R},{\hat
  R}) = {\hat R} + g({\bar R})$ is equivalent to a theory of the type
(\ref{eq:action0}) with $f({\bar R}) = {\bar R} + g({\bar R})$.  The choice
$f({\bar R},{\hat R}) = {\bar R} + g({\hat R})$ can be obtained from
the above analysis with $\varepsilon_1=1$, $\rho = \varphi = \Psi=0$.
In this case the formula (\ref{eq:Vdef}) for the potential becomes $$V(\Phi) =
\frac{e^{4\alpha(\Phi)}} {2 \kappa^2} \left[ \psi g'(\psi) - g(\psi)
  \right]$$ where $\tanh^2(\kappa \Phi/\sqrt{6}) = |g'(\psi)|$.

\section{Discussion and implications}

Consider first the compatibility of the theory (\ref{eq:action4}) with
solar system 
experiments.  Lets assume that the background, present-day
cosmological values of the fields $\Phi$ and $\Psi$ are $\Phi_0$ and
$\Psi_0$.  Lets also assume, initially, that the potential $V$ is such
that the fields $\delta \Phi = \Phi - \Phi_0$ and $\delta \Psi = \Psi
- \Psi_0$ have small masses and are long ranged on solar
system scales.
The Eddington parameterized post-Newtonian (PPN) parameter
$\gamma$ for theory is then given by
\cite{Damour92} 
\begin{equation}
1 - \gamma = \frac{2 \alpha_0^2}{1 + \alpha_0^2},
\label{eq:gammaformula}
\end{equation}
where
\begin{equation}
\alpha_0^2 = \frac{2}{\kappa^2} \gamma^{AB} \frac{\partial
  \alpha}{\partial \Phi^A} \frac{\partial \alpha}{\partial \Phi^B}
\label{eq:alpha0def}
\end{equation}
evaluated at the present-day background values of the fields.
Evaluating the expression (\ref{eq:gammaformula}) using Eqs.\
(\ref{eq:sigmametricI}) -- 
(\ref{eq:alphadefI}) gives $\gamma = 1/2$.
This is in conflict with solar system VLBI observations
which show that \cite{Will}
\begin{equation}
|\gamma -1| \le 3 \times 10^{-4}.
\label{eq:expt}
\end{equation}
It follows that the potential $V(\Phi,\Psi)$ must be such that at least
one of the two mass eigenstates for perturbations about $(\Phi_0,\Psi_0)$
is short ranged.  Those mass eigenstates are determined from the
equations \cite{Damour92}
\begin{eqnarray}
\gamma_{AB}(\Phi_0^C) &=& \sum_j \lambda_A^j \lambda_B^j  \\
\nabla_A \nabla_B V(\Phi_0^C) &=& \sum_j m_j^2 \lambda_A^j \lambda_B^j,
\end{eqnarray}
where $j$ labels the eigenstate, $m_j$ are the masses, and $\nabla_A$
is the covariant 
derivative determined by the metric $\gamma_{AB}$ on field space.  
The formula (\ref{eq:alpha0def}) is then replaced by
\begin{equation}
\alpha_0^2 = \frac{2}{\kappa^2} \sum_j^\prime \left( \gamma^{AB}
  \lambda^j_A \frac{\partial \alpha}{\partial \Phi^B} \right)^2,
\label{eq:alpha0defa}
\end{equation}
where $\sum_j^\prime$ denotes the sum over the light, long-ranged
eigenstates.

If there is one light and one heavy eigenstate, and if $\chi$ is the
angle in field space between the light eigenstate and the $\partial /
\partial \Phi$ direction, then a short calculation 
using Eqs.\ (\ref{eq:sigmametricI}), (\ref{eq:alphadefI}) and
(\ref{eq:alpha0defa}) gives 
\begin{equation}
\alpha_0^2 = \frac{ \left[ \sin \chi + \cos \chi \sinh (\kappa \Phi_0 /
    \sqrt{6}) \right]^2 }{3 \cosh^2 (\kappa \Phi_0 / \sqrt{6})}.
\end{equation}
Thus if $\chi \ll 1$ and $\kappa \Phi_0 \ll 1$, then $\alpha_0^2 \ll
1$ and the solar system constraint (\ref{eq:expt}) can be evaded.  It
is possible to find functions $f({\bar R}, {\hat R})$ and values of
$(\Phi_0$, $\Psi_0)$ for which $\chi \ll 1$.  This is because $\chi$ is
determined by the values of $f$ and its derivatives up to third order
at the point $({\bar R}_0,{\hat R}_0)$ that corresponds to
$(\Phi_0,\Psi_0)$, so that there many more free parameters than there
are constraints.  Thus, there exists a class of gravitation
theories of this type that are compatible with solar system observations.

It might be possible to further specialize the choice of $f({\bar
  R},{\hat R})$ to 
give a viable model of the recent acceleration of the Universe, along
the lines suggested by Carroll {\it et. al.} \cite{Carroll}.  However,
in such a case, it would not be completely accurate to call the theory
(\ref{eq:action3}) a modified theory of gravity, since 
the theory would contain a scalar field which acts mostly as a source
of gravity and does not mediate any long range forces.  Rather, the
explanation for the acceleration of the Universe would in such a case
be partly a modification of gravity and partly a quintessence-type
source of gravity.

\medskip

\acknowledgments
I thank Nima Arkani-Hamed and Jim Cline for helpful discussions.
This research was supported in part by NSF grant PHY-0140209.

\end{document}